\documentclass{SciPost}

\hypersetup{
    colorlinks,
    linkcolor={red!50!black},
    citecolor={blue!50!black},
    urlcolor={blue!80!black}
}

\usepackage[bitstream-charter]{mathdesign}
\DeclareSymbolFont{usualmathcal}{OMS}{cmsy}{m}{n}
\DeclareSymbolFontAlphabet{\mathcal}{usualmathcal}

\fancypagestyle{SPstyle}{
\fancyhf{}
\lhead{\colorbox{scipostblue}{\bf \color{white} ~SciPost Physics Proceedings }}
\rhead{{\bf \color{scipostdeepblue} ~Submission }}

\fancyfoot[C]{\textbf{\thepage}}
}

\begin{document}

\pagestyle{SPstyle}

\begin{center}{\Large \textbf{\color{scipostdeepblue}{QGSJET-III: predictions 
for extensive air shower characteristics and the corresponding 
uncertainties}}}\end{center}

\begin{center}\textbf{
Sergey Ostapchenko\textsuperscript{$\star$}
}\end{center}

\begin{center}
Universit\"at Hamburg, II Institut f\"ur Theoretische
Physik, 22761 Hamburg, Germany
\\[\baselineskip]
$\star$ \href{mailto:email1}{\small sergey.ostapchenko@desy.de}
\end{center}

\definecolor{palegray}{gray}{0.95}
\begin{center}
\colorbox{palegray}{
  \begin{tabular}{rr}
  \begin{minipage}{0.36\textwidth}
    \includegraphics[width=55mm]{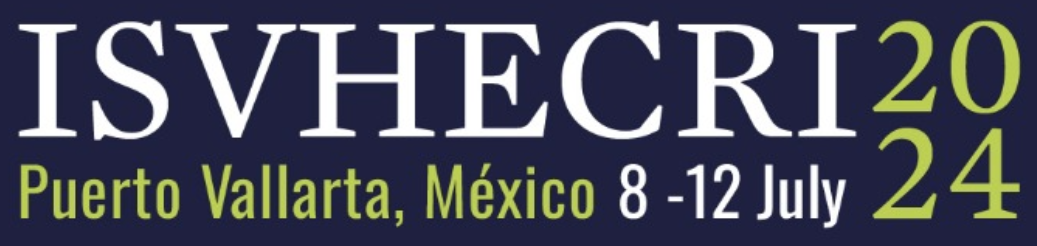}
  \end{minipage}
  &
  \begin{minipage}{0.55\textwidth}
    \begin{center} \hspace{5pt}
    {\it 22nd International Symposium on Very High \\Energy Cosmic Ray Interactions (ISVHECRI 2024)} \\
    {\it Puerto Vallarta, Mexico, 8-12 July 2024} \\
    \doi{10.21468/SciPostPhysProc.?}\\
    \end{center}
  \end{minipage}
\end{tabular}
}
\end{center}

\section*{\color{scipostdeepblue}{Abstract}}
\textbf{\boldmath{The physics content of the QGSJET-III Monte Carlo model of high energy hadronic interactions is briefly described. The predictions of the model for   extensive air shower characteristics  are presented in comparison to the corresponding results of other  Monte Carlo generators of cosmic ray interactions. The results of a recent quantitative analysis of uncertainties for such predictions are discussed, notably, regarding possibilities to enhance the muon content of   extensive air showers or to delay the air shower development.}}

\vspace{\baselineskip}

\noindent\textcolor{white!90!black}{%
\fbox{\parbox{0.975\linewidth}{%
\textcolor{white!40!black}{\begin{tabular}{lr}%
  \begin{minipage}{0.6\textwidth}%
    {\small Copyright attribution to authors. \newline
    This work is a submission to SciPost Phys. Proc. \newline
    License information to appear upon publication. \newline
    Publication information to appear upon publication.}
  \end{minipage} & \begin{minipage}{0.4\textwidth}
    {\small Received Date \newline Accepted Date \newline Published Date}%
  \end{minipage}
\end{tabular}}
}}
}


\vspace{10pt}
\noindent\rule{\textwidth}{1pt}
\tableofcontents
\noindent\rule{\textwidth}{1pt}
\vspace{10pt}


\section{Introduction}
\label{sec:intro}
Experimental studies of  ultra-high energy cosmic rays (UHECRs) are traditionally performed using indirect methods: based on measurements
of the so-called  extensive air showers (EAS) --
nuclear-electromagnetic cascades initiated by interactions of UHECRs in the
atmosphere \cite{nag00}. Therefore, an analysis and interpretation of the
corresponding experimental data requires an accurate 
  description of EAS development, notably, regarding the
cascade of nuclear interactions of both primary cosmic ray (CR) 
particles and of secondary hadrons produced. Here comes the importance of Monte Carlo (MC) generators of hadronic interactions, employed 
in EAS simulation procedures \cite{eng11}. In turn, since such MC  generators are largely phenomenological and involve a considerable extrapolation of the underlying physics into scarcely studied kinematic regimes, of considerable importance is to estimate the range of uncertainty regarding EAS predictions of such models \cite{ulr11,par11}.

 In the current contribution, the results of recent quantitative investigations of model uncertainties for EAS predictions \cite{ost24a,ost24b}, in the framework of the QGSJET-III model  \cite{ost24c,ost24d}, are discussed. The corresponding studies were guided by  three basic principles:  i) the changes of the corresponding modeling were performed at a 
microscopic level; ii) the considered modifications were restricted by the requirement
not to contradict basic physics principles; iii) the consequences of such changes,
regarding a potential (dis)agreement with relevant accelerator data, were analyzed.

\section{QGSJET-III model}
\label{sec:q-iii}
The major development in QGSJET-III, compared to the previous model version,
QGSJET-II-04 \cite{ost11,ost13}, concerns the treatment of nonlinear corrections to perturbative hard scattering processes \cite{ost24c}. The corresponding standard approach
in all present MC generators of hadronic collisions is based on the leading twist  collinear factorization of perturbative quantum chromodynamics (pQCD) \cite{col89}.
 In that case, the inclusive parton jet production cross section is defined by a convolution of two parton momentum distribution functions (PDFs) of interacting hadrons (nuclei) with the Born parton scatter cross section, thus corresponding to a binary parton-parton scattering. However, since such a cross section explodes in the limit of small jet transverse momentum $p_t$, in MC models one is forced to introduce a low  $p_t$ cutoff for jet production  and all the model
predictions depend strongly on the choice of that cutoff.

In QGSJET-III, one considered a phenomenological implementation  
of a certain class of   higher twist corrections to  hard parton-parton
scattering, namely, those which  correspond to multiple coherent rescattering
of final $s$-channel partons on correlated ``soft'' gluon pairs, characterized
by very small light cone (LC) momentum fractions $x$ \cite{qiu04,qiu06}.
In such a case, the hardest scattering process is no longer a binary parton-parton scattering but generally involves an arbitrary number of soft gluons. As demonstrated in \cite{ost19}, such a development reduces considerably the dependence of model predictions on the  low  $p_t$ cutoff.

An additional technical improvement in the QGSJET-III model concerned a more
consistent treatment of the pion exchange process in hadronic collisions \cite{ost21,ost24d}, including a cross check of the approach, based on the data 
of the LHCf experiment on forward neutron production in proton-proton interactions \cite{adr15,adr18}. As demonstrated earlier in \cite{ost13}, assuming a dominance
of the $t$-channel pion exchange in pion-nucleus collisions, over  
contributions of heavier  Reggeon states (e.g., of $\rho$-mesons), gives rise to a substantial
($\simeq 20$\%) enhancement of EAS muon content $N_{\mu}$, due to a significant
increase of forward $\rho$-meson production. This is somewhat nontrivial, being a direct consequence of the isospin symmetry.\footnote{While the isospin symmetry is not an exact one for strong
interactions, it holds to a very good accuracy thanks to the small mass difference between the $u$ and $d$ quarks.}
Enhancing central production of  $\rho$-mesons, at an expanse of pions,
would not have an appreciable impact on the predicted  $N_{\mu}$ (e.g., \cite{ost24a}): since  $\rho ^+$,  $\rho ^-$, and $\rho ^0$ are created in proportion
1:1:1 and, upon decays of those mesons ($\rho^{\pm}\rightarrow \pi^{\pm}\pi^0$,
$\rho^{0}\rightarrow \pi^{+}\pi^-$), the energy partition between the resulting
charged and neutral pions is the same as for direct pion production
($E_{\pi^{\pm}}$:$E_{\pi^{0}}$=2:1). On the contrary, for   $\rho$-mesons resulting
from the pion exchange process, this energy partition becomes 3:1
($\pi^{\pm}\xrightarrow[\pi^{0}]{}\rho^{\pm}\rightarrow\pi^{\pm}\pi^{0}$,
$\pi^{\pm}\xrightarrow[\pi^{\pm}]{}\rho^{0}\rightarrow\pi^{+}\pi^{-}$).
Consequently, a higher pion exchange rate leads to a larger fraction of the primary particle energy, 
  retained in the nuclear cascade at a given depth, 
   instead of going into the electromagnetic ``sink''
via the $\pi^0\rightarrow \gamma\gamma$ decay, thereby giving rise to a higher $N_{\mu}$.

\begin{figure}[t]
\centering
\includegraphics[width=0.44\textwidth]{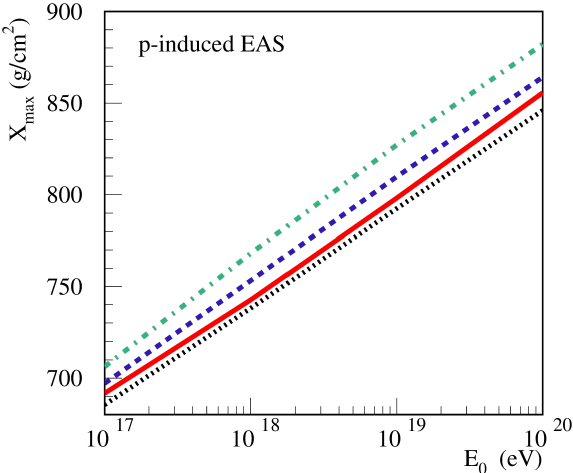}\hfill
\includegraphics[width=0.44\textwidth]{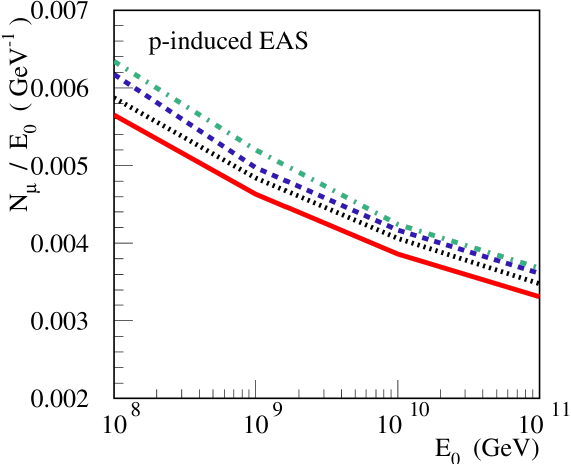}
\caption{Dependence on primary energy of the shower maximum depth $X_{\max}$ (left)
and   of the muon number  $N_{\mu}$   at sea level (right),
for proton-initiated EAS,
calculated using the 
QGSJET-III, QGSJET-II-04, EPOS-LHC, and SIBYLL-2.3 models --
solid, dotted, dashed, and dash-dotted lines, respectively.
}
\label{fig:xmax}
\end{figure}

Regarding the model predictions for  basic EAS characteristics, those appeared 
to be rather similar to the ones  of the previous model version, QGSJET-II-04:
the  difference for the predicted extensive air  shower maximum depth $X_{\max}$
being $\leq 10$ g/cm$^2$, while the one for EAS muon number  $N_{\mu}$
($E_{\mu}>1$ GeV) amounting to 5\% only. This is illustrated in  Fig.~\ref{fig:xmax},
where the corresponding results of the two models are compared to each other
and to predictions of two other CR interaction models, EPOS-LHC \cite{pie15}
and SIBYLL-2.3d \cite{rie20}. Such a robustness of the calculated EAS characteristics
may suggest that the relevant features of interaction models are 
sufficiently constrained by accelerator data.\footnote{Potential explanations for the somewhat different results
of the  EPOS-LHC and SIBYLL-2.3d models have been discussed in \cite{ost23}.}

\section{Uncertainties for the predicted EAS muon content}
\label{sec:nmu}

One of the traditional methods for high energy CR composition studies is based
on measurements of ground lateral density of muons in extensive air showers \cite{nag00,kam12}. However, the use of this method for UHECRs is hampered presently by a
persisting contradiction between the corresponding predictions of EAS simulations and the experimental data, the latter indicating a substantially higher EAS muon content \cite{aab15,aab16}.

Generally, the predicted $N_{\mu}$ is correlated with the multiplicity of hadron-air collisions (e.g., \cite{ulr11}), which can be understood  using the simple Heitler's qualitative picture for the cascade process \cite{hei54,mat05}.
 Yet the
relation between the multiplicity and the shower muon size  is not straightforward since more energetic secondary hadrons capable of producing 
powerful enough subcascades give larger contributions to  $N_{\mu}$, compared
to much more copiously produced low energy hadrons. As demonstrated, e.g., 
in \cite{ost24a}, the competition between an abundant production of low energy hadrons and larger muon yields from high energy secondaries leads to $N_{\mu}$ 
being approximately proportional  to the quantity  
$\langle x_E^{\alpha_{\mu}} n_{\rm stable}^{\pi -{\rm air}}\rangle$,  defined as
\begin{equation}
 \langle x_E^{\alpha_{\mu}}\, n_{\rm stable}^{\pi -{\rm air}}(E_0)\rangle =  \int \!dx_E\; x_E^{\alpha_{\mu}}\,
 \frac{dn_{\rm stable}^{\pi -{\rm air}}(E_0,x_E)}{dx_E}\,, 
 \label{x*N.eq}
\end{equation}
where $dn_{\rm stable}^{\pi -{\rm air}}/dx_E$ is the distribution, with respect to
 the energy fraction $x_E$, of   ``stable'' secondary hadrons in pion-air collisions,
  i.e., those which have significant chances to interact in the atmosphere, instead of decaying. 
In turn, $\alpha_{\mu}\simeq 0.9$ is
the characteristic exponent  for the dependence of $N_{\mu}$ on the primary
energy, for proton-induced EAS: $N_{\mu}^p(E_0)\propto E_0^{\alpha_{\mu}}$.
Since  $\alpha_{\mu}$ is not too different from unity, the quantity
 $\langle x_E^{\alpha_{\mu}} n_{\rm stable}^{\pi -{\rm air}}\rangle$
can be approximated by the average fraction of the parent pion energy 
$\langle x_E n_{\rm stable}^{\pi -{\rm air}}\rangle$ (the case
 $\alpha_{\mu}=1$), taken by all stable secondary hadrons. Therefore, 
 to predict a higher EAS muon content, a higher energy fraction taken by all
 stable secondaries in pion-air interactions is required.
 
 Since experimental data on pion-proton and pion-nucleus collisions are available
 at fixed target energies only, one may try to enlarge the predicted  $N_{\mu}$ by
 enhancing the energy-rise of secondary hadron yields. The corresponding energy
 dependence is driven by (mini)jet production, which is, in turn, governed by gluon
 PDFs. Hence, the simplest way to obtain the desirable enhancement is to change the
 LC momentum partition between valence quarks and gluons (plus sea quarks) in the 
 pion, in favor of the latter. It is worth remarking, however, that valence quark
 PDFs of the pion are seriously constrained by experimental studies of the Drell-Yan
 process in pion-proton scattering, while similar constraints on gluon PDFs come
 from measurements of direct photon and $J$/$\psi$ production. Neglecting for the moment
 those constraints, one may try an extreme scenario: reducing the pion LC momentum
 fraction carried by valence quarks by factor  two and enhancing
 correspondingly the gluon content of the pion. Yet such an extreme modification allows
 one to enlarge $N_{\mu}$ by less than 1\% \cite{ost24a}. 
This is because a noticeable enhancement of secondary hadron yields
is obtained this way only in central rapidity 
 region, i.e., for small $x_E$, and for sufficiently high pion energies corresponding
 to the top part of the nuclear cascade in the atmosphere.
 
\begin{figure}[t]
\centering
\hspace{1cm}
\includegraphics[width=0.16\textwidth]{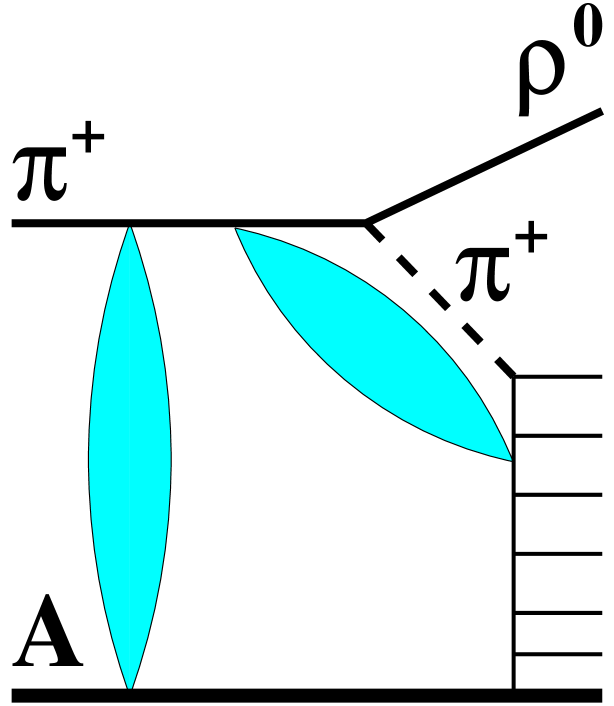}
\hspace{1cm}
\includegraphics[width=0.16\textwidth]{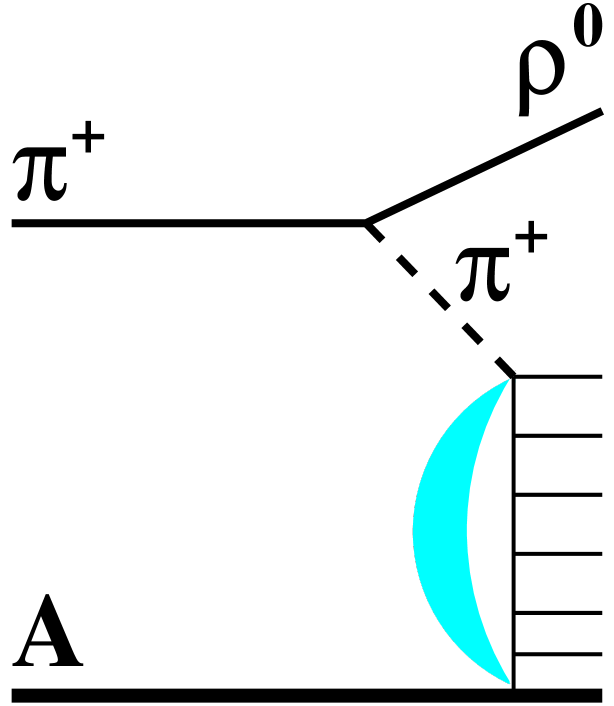}
\hspace{1cm}
\includegraphics[width=0.16\textwidth]{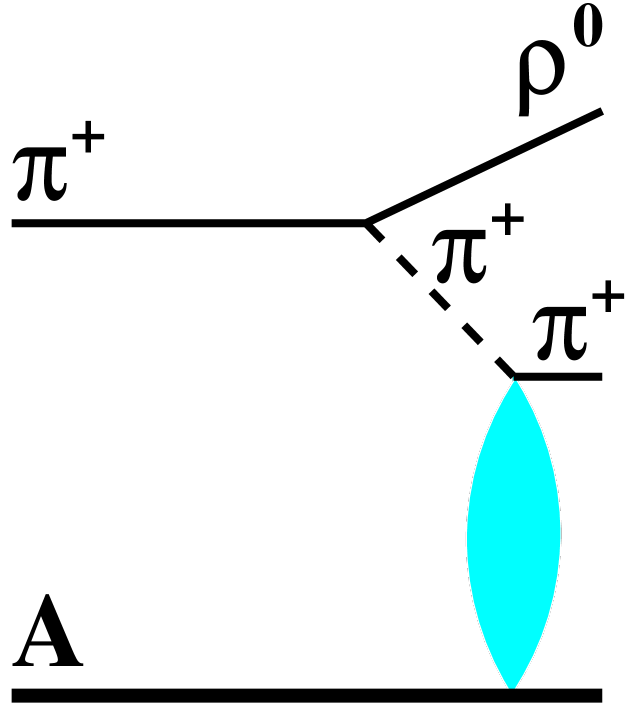}
\hspace{1cm}
\caption{Left: schematic view of the pion exchange process in pion-nucleus interaction;
shown by light-shaded ellipses are absorptive corrections due to additional rescatterings
of the incident pion. Neglecting such corrections, one has to account both for the
inelastic (middle) and elastic (right) interaction of the virtual pion.}
\label{fig:piex}
\end{figure}

 Further, in view of the strong impact of the pion exchange process on the predicted
  $N_{\mu}$, one may try to change the energy dependence of that process. This dependence 
  is governed in QGSJET-III by the so-called absorptive corrections, i.e., by the 
  probability not to have additional inelastic rescatterings in pion-air interactions,
  see Fig.~\ref{fig:piex}~(left). Indeed, such additional rescatterings would ``suck out''
  energy from the pion, thereby preventing a production of $\rho$-mesons with large  $x_E$.
  Because of the general energy-rise of multiple scattering, the absorptive corrections
  ``push'' the pion exchange process towards larger and larger impact parameters, with
  increasing energy, giving rise to a slow decrease of the pion exchange rate \cite{ost23}.
Yet here one has some space for a model dependence, e.g., if additional inelastic
 rescatterings take only small energy fractions, like in the SIBYLL model (see  
 the corresponding discussion in  \cite{ost16}). Therefore, one may try an extreme modification:
 neglecting such absorptive corrections completely and having thus an energy-independent
 probability for the pion exchange process. Paradoxically, such a change would result in
 a decrease of the predicted  $N_{\mu}$ (by up to 10\% at $E_0=10^{19}$ eV) \cite{ost24a}.
 In the absence of absorptive corrections, in addition to the inelastic interaction of the
 virtual pion with the target nucleus, shown in  Fig.~\ref{fig:piex}~(middle), 
 one has
 to take into account the contribution of pion elastic scattering of  Fig.~\ref{fig:piex}~(right).
 It is the scarce hadron production in the latter case which causes the decrease of 
  $N_{\mu}$ \cite{ost24a}.
  
\begin{figure}[t]
\centering
\includegraphics[width=0.44\textwidth]{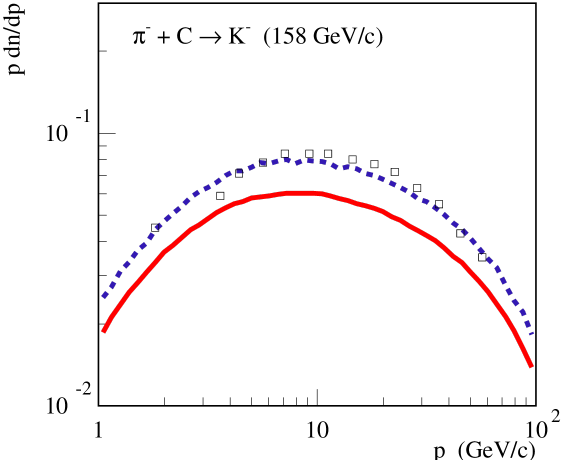}\hfill
\includegraphics[width=0.44\textwidth]{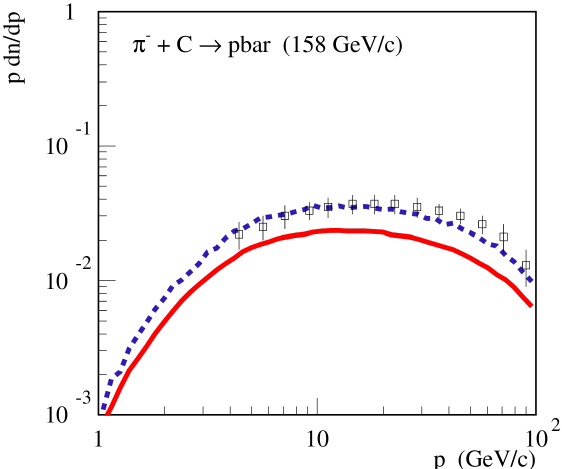}
\caption{Momentum distributions in laboratory frame of $K^-$ (left) and of $\bar p$ (right) 
 produced in $\pi^-$C collisions at 158 GeV/c,
   calculated using the default QGSJET-III model
(solid lines) or considering 40\% and 60\% enhancement of kaon and (anti)nucleon yields,
respectively, compared to NA61 data \cite{adh23} (points).}
\label{fig:na61}
\end{figure}

\begin{figure}[t]
\centering
\includegraphics[width=0.44\textwidth]{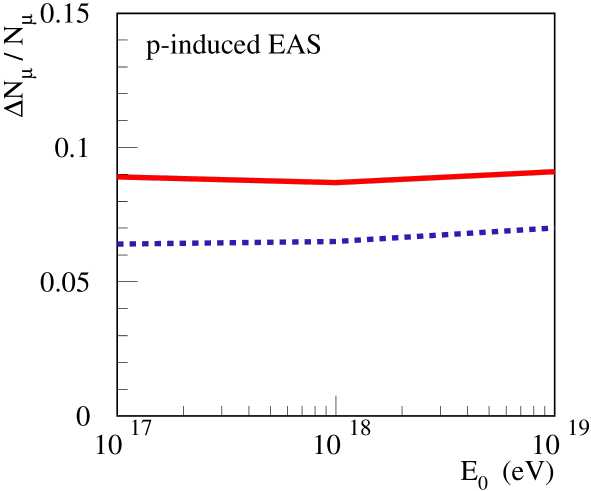}
\caption{Dependence on primary energy of the 
relative change of  the muon number $N_{\mu}$ at sea level ($E_{\mu}>1$ GeV), for 
proton-initiated air shower,
for  60\% enhancement of (anti)nucleon production (solid line) and for 40\%
enhancement of   kaon  production (dashed line).}
\label{fig:nmu-enh}
\end{figure}

  Thus, the only viable option for enlarging significantly the predicted EAS muon content
  is to enhance relative yields of secondary kaons and (anti)nucleons in pion-air collisions,
  at the expanse of pions -- since this would decrease the energy leak into neutral pions.
  Comparing in Fig.~\ref{fig:na61} the corresponding results of the  QGSJET-III model
   with the data of the NA61 experiment, we see that a significant enhancement of  the 
   predicted yields is required to match the data: $\simeq 40$\% for kaons and  $\simeq 60$\%
   for (anti)nucleons.\footnote{However, as discussed in \cite{ost24a}, such modifications would
   lead to a serious tension with results of other experiments on kaon and (anti)proton
   production in $pp$ and $\pi p$ collisions.} Applying
   such changes allows one to enhance the  predicted $N_{\mu}$ by up to 10\%,
    see Fig.~\ref{fig:nmu-enh} \cite{ost24a}.

\section{Uncertainties for the predicted EAS maximum depth}
\label{sec:xmax}

Let us now turn to the EAS maximum depth $X_{\max}$ which is the main air shower
characteristic used for UHECR composition studies \cite{kam12}. Unlike the EAS muon
content which depends on the whole history of the nuclear cascade in the atmosphere,  $X_{\max}$ is largely governed by interactions of primary CR particles. Consequently, the corresponding model results are seriously constrained by experimental data of the Large Hadron Collider (LHC).
Nonetheless, there exist some tension between the predictions of EAS simulations for $X_{\max}$ and the data of the Pierre Auger Observatory: as demonstrated in \cite{pao24}, to reach a consistency with the measurements, a significantly slower air shower development is required.

What are the possibilities to have a larger $X_{\max}$ predicted? First of all,
a smaller $pp$ inelastic cross section, $\sigma^{\rm inel}_{pp}$, would correspond to a smaller cross section for proton collisions with air, 
 $\sigma^{\rm inel}_{p-{\rm air}}$, in the Glauber-Gribov formalism \cite{gla59,gri69}.
In turn, this would enlarge the proton mean free path in air, 
$\lambda_p\propto 1/\sigma^{\rm inel}_{p-{\rm air}}$, thereby shifting the
whole air shower profile towards larger depths. A similar effect can be obtained
by increasing the rate of diffractive interactions, 
$\sigma^{\rm diffr}_{p-{\rm air}}/\sigma^{\rm inel}_{p-{\rm air}}$,
with $\sigma^{\rm diffr}_{p-{\rm air}}$ being the cross section for inelastic 
diffraction on air. Indeed, since in diffractive collisions the proton looses typically a small portion of its energy, the effect of diffraction on EAS development is more or less equivalent to a redefinition of the  proton mean free path: $\lambda_p\rightarrow \lambda_p (1+\sigma^{\rm diffr}_{p-{\rm air}}/
\sigma^{\rm inel}_{p-{\rm air}})$.

\begin{figure}[t]
\centering
\includegraphics[width=0.44\textwidth]{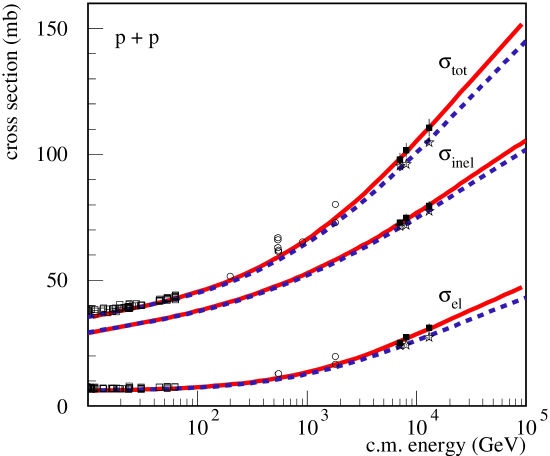}\hfill
\includegraphics[width=0.44\textwidth]{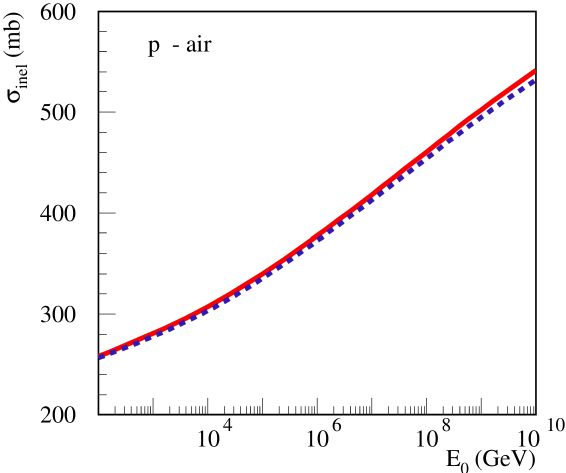}
\caption{Center of mass (c.m.)  energy dependence of the total, inelastic, and
 elastic $pp$ cross sections, compared to experimental data \cite{pdg,ant19,aad23}  (left) and   
laboratory  energy dependence of   $\sigma^{\rm inel}_{p-{\rm air}}$
(right),  
calculated  with the default QGSJET-III model
(solid lines) and considering a 30\% enhancement of low mass  diffraction 
 (dashed lines).}
\label{fig:sigpp}
\end{figure}

In the QGSJET-III model, one can study the combined effect of both, a smaller
 $\sigma^{\rm inel}_{p-{\rm air}}$ and a larger $\sigma^{\rm diffr}_{p-{\rm air}}$, by increasing the cross section for low mass diffraction in $pp$ collisions \cite{ost24b}. Enhancing the low mass ($\leq 3.4$ GeV) in $pp$ by $\simeq 30$\% and being still compatible with the corresponding results of the TOTEM experiment \cite{ant13a}, one obtains $\simeq 15$\% higher rate of diffractive-like proton-air interactions characterized by a small ($<10$\%) energy loss of leading nucleons. On the other hand, since a higher diffraction is bound to a stronger inelastic screening effect \cite{gri69}, this leads to  smaller total, inelastic, and elastic proton-proton cross sections, as illustrated in Fig.\ \ref{fig:sigpp}~(left), all becoming compatible with the data of the ATLAS experiment,
 shown by the open stars in the Figure.
Yet the corresponding reduction of $\sigma^{\rm inel}_{p-{\rm air}}$ is $\leq 1$\%,   see Fig.\ \ref{fig:sigpp}~(right), since a proton-nucleus cross section is largely dominated by the nuclear size. Therefore, the effect of the considered changes on  $X_{\max}$ is largely caused by the enhanced diffraction rate in proton-air interactions. 
The obtained shift of the average EAS maximum depth is limited by $\simeq 8$ g/cm$^2$ \cite{ost24b},
 in a good agreement with earlier studies \cite{ost14}.

Another possibility to obtain a larger  $X_{\max}$ predicted is to slow down the energy rise
of the inelasticity  $K^{\rm inel}_{p-{\rm air}}$ of proton-air interactions:
since this would enlarge somewhat the average number of hadron ``generations''
in the nuclear cascade, thereby elongating the air shower profile.
The energy rise of the inelasticity is a generic feature: since the rate of multiple
scattering in hadronic collisions increases with energy.
 However, the speed of predicted energy rise of   $K^{\rm inel}_{p-{\rm air}}$ may 
vary from  model to model, depending, e.g., on how much energy is taken
by a single inelastic rescattering process \cite{ost16}.
 Since the energy dependence of multiple scattering rate is driven by a fast
 increase of (mini)jet production, one may try to tame that rise by considering
  stronger higher twist effects in the QGSJET-III model, which would further
 suppress the emission of minijets of relatively small transverse momenta \cite{ost24b}.
 On the other hand, the increase of the inelasticity would be reduced 
  if additional inelastic
 rescatterings took only small portions of  energy of the incident proton.
 This can be achieved by  choosing a softer  distribution,
 $\propto x^{-\alpha_{\rm sea}}$, for the LC momentum fraction $x$ of constituent
   sea (anti)quarks
involved in such rescattering processes \cite{par11,ost16}, i.e., using a larger
value for  $\alpha_{\rm sea}$.
  
\begin{figure}[t]
\centering
\includegraphics[width=0.44\textwidth]{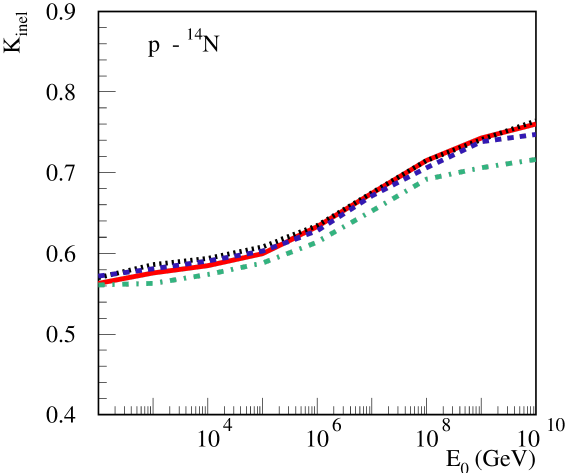}\hfill
\includegraphics[width=0.44\textwidth]{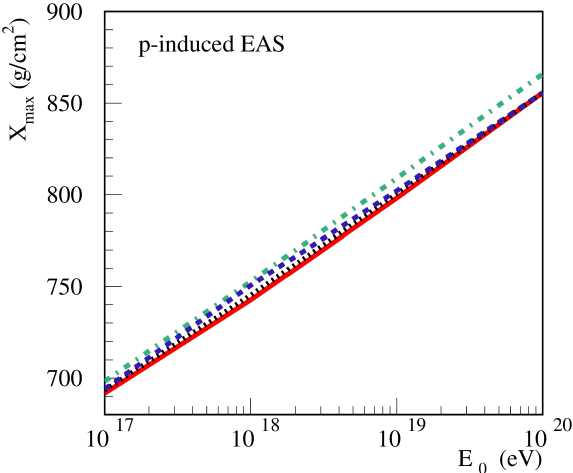}
\caption{Laboratory energy dependence  of    $K^{\rm inel}_{pN}$ (left) and 
 primary energy dependence of $X_{\max}$ for $p$-induced EAS (right),
 for  the default QGSJET-III model (solid line) and for the model
 modifications discussed in the text;  dotted,  dashed, and 
  dash-dotted lines correspond to  $\alpha_{\rm sea}=0.65$, 0.8, and 0.9,
  respectively.}
\label{fig:xmax-hts}
\end{figure}

Considering twice  stronger higher twist effects, using 
$\alpha_{\rm sea}=0.8$ and $\alpha_{\rm sea}=0.9$, in addition to the
default value $\alpha_{\rm sea}=0.65$, and
 adjusting the parameters of the hadronization procedure of the QGSJET-III  model
in order to keep an agreement with accelerator data,
one arrives  to the energy dependence of the inelasticity of proton-nitrogen
interactions, $K^{\rm inel}_{pN}$,  shown in Fig.\ \ref{fig:xmax-hts}~(left), 
while the corresponding results
for EAS maximum depth are plotted in Fig.\ \ref{fig:xmax-hts}~(right).
As one can see in Fig.\  \ref{fig:xmax-hts}, noticeable changes both for 
$K^{\rm inel}_{pN}$ and  $X_{\max}$ are caused only by modifications of LC
momentum distributions of constituent partons. In particular, using
  $\alpha_{\rm sea}=0.9$, one obtains  up to $\simeq 6$\% reduction of 
   $K^{\rm inel}_{pN}$ and up to $\simeq 12$ g/cm$^2$ larger  $X_{\max}$
  at the highest energies \cite{ost24b}.

Generally, one could have expected a stronger dependence of the inelasticity 
and of the predicted EAS maximum depth on the 
momentum distributions of constituent partons \cite{ost03}. However, at very high
energies, this naive picture is substantially modified due to the dominance
of (semi)hard scattering processes over purely nonperturbative soft
interactions. Indeed,
in any partial semihard rescattering, the hardest  parton-parton scattering 
 is usually preceded by multiple emission of ``softer'' partons (so-called
 initial state radiation): with each ``parent'' parton in the cascade
 having a higher momentum fraction than its ``daughter''. 
Hence, the total momentum fraction taken by the
 first $s$-channel partons produced in such
perturbative   cascades constitutes a   lower bound on the inelasticity.

In principle,  one may consider a more exotic scenario, assuming that the standard
hadron production  pattern  is significantly modified at very high 
energies by collective effects. While this could allow one to 
 increase the predicted  $X_{\max}$ more significantly,
  by up to  $\simeq 30$ g/cm$^2$, such modifications
are seriously disfavored both  by the data of the LHCf experiment \cite{adr15,adr18}, 
regarding forward neutron production in $pp$ collisions at LHC,
and   by measurements of the  muon production depth
 at the Pierre Auger Observatory
 \cite{pao-nmu}, as demonstrated in  \cite{ost24b}.

\section{Conclusion}
\label{sec:Outlook}
In this contribution, I briefly discussed the physics content of the QGSJET-III  model
and presented its  predictions for basic  extensive air showers characteristics.
Overall, the EAS results of  QGSJET-III are rather close to the ones of the QGSJET-II-04
model: with the predicted  $X_{\max}$ being up to 10 g/cm$^2$ larger and with $N_{\mu}$
being reduced by $\simeq 5$\%. Such a robustness of the model predictions for EAS
development is not occasional but rather reflects the fact that the relevant features
of high energy interaction treatment are sufficiently constrained by
 accelerator measurements, notably, from the Large Hadron Collider.
Indeed, performing a quantitative analysis of the corresponding model uncertainties,
within the standard physics picture and within the limits allowed by available accelerator
data, one was able to further increase the predicted  $X_{\max}$ by up to $\simeq 10$ g/cm$^2$ only,
while the predicted EAS muon content could be enhanced by $\simeq 10$\%  \cite{ost24a,ost24b}.

\paragraph{Funding information}
This work was supported by  Deutsche
 Forschungsgemeinschaft (project number 465275045).







\bibliography{ostapchenko_isvhecri2024.bib}


\end{document}